# Is response priming based on surface color? Response to Skrzypulec (2021)

## Thomas Schmidt

Thomas Schmidt
University of Kaiserslautern, Germany
Experimental Psychology Unit
www.sowi.uni-kl.de/psychologie
thomas.schmidt@sowi.uni-kl.de

### Abstract

**Skrzypulec (2021) raises the question whether motor activation by masked color primes is based on the same type of color representation as conscious vision. He postulates that the literature on color processing without awareness makes an implicit assumption that "conscious" and "unconscious" color representations have the same properties, in which case priming by masked color stimuli would indeed indicate that the same complex representation of surface color can be conscious as well as unconscious. I review some evidence from response priming by lightness stimuli in the context of a visual illusion that alters the perceived lightness of the primes (Schmidt et al., 2010). Those results clearly show that response priming is not driven by color-constant information but instead by local image contrast, making it unlikely that rapid response activation by color primes is based on a color-constant representation of surface color.**

Skrzypulec (2021) reviews the literature on color processing without awareness in different paradigms, one of them being response priming by masked colors (Breitmeyer, Ro, Öğmen, & Todd, 2007; Breitmeyer, Ro, & Singhal, 2004; Brenner & Smeets, 2004; Ro, Singhal, Breitmeyer, & Garcia, 2009; Schmidt, 2000, 2002; Tapia, Breitmeyer, & Shooner, 2010). In a typical experiment, participants respond to one of two color targets (e.g., a red or green annulus) preceded by one of two color primes (e.g., a red or green dot that briefly appears at the same position as the target and that exactly fills its inner contour, which leads to *metacontrast masking* of the prime). When primes and targets are *consistent* (mapped to the same response), responses to the targets are fast and error rates are low. When they are *inconsistent* (mapped to different responses), responses to the target are slower and response errors appear. These priming effects in response times and error rates increase with increasing stimulus-onset asynchrony (SOA) between prime and target. Response priming by color has also been studied in the time-course of pointing movements (Schmidt, 2002; Schmidt, Niehaus, & Nagel, 2006; Schmidt & Schmidt, 2010; Schmidt & Seydell, 2008) and lateralized readiness potentials (Vath & Schmidt, 2007). While many studies employ backward masking by metacontrast to reduce the visibility of the prime, other studies employ flanker designs where the prime remains unmasked (Wolkersdorfer, Panis, & Schmidt, 2020). Biafora and Schmidt (2020) showed that response priming effects can increase with SOA even though prime discrimination performance decreases, thus demonstrating a double dissociation between the two measures of color processing (Schmidt & Vorberg, 2006).

Skrzypulec's major concern is the following: Even if we accept that response priming by color occurs under complete masking of the prime, can we conclude that it is based on a representation similar to conscious color vision? Skrzypulec particularly distinguishes between representations of *surface color* and *relational color*. Surface color is characterized by *color constancy*, the ability to perceive an invariance in the reflectance spectrum[1] of a colored surface even in the presence of differences in illumination (e.g., by shadows, colored illuminants, or an intervening transparent medium through which the surface is viewed). Relational color, as I would explain it, is the minimum information needed to distinguish between the two stimuli in the experiment, e.g., a red prime from a green prime. Such a representation need not be phenomenally rich, or similar to conscious perception, or color-constant; it only needs to serve the purpose of mapping stimuli to the correct responses. Skrzypulec believes that research on unconscious color processing is based on the strong implicit assumption that unconscious color representations are actually the same as the conscious ones, only devoid of phenomenal awareness of color. If this *equivalence hypothesis* were true, priming by masked color stimuli would indeed indicate that the same complex representation of surface color (including color constancy) can be conscious as well as unconscious.

I find the idea of identical representations in visuomotor response priming and conscious color vision quite odd. The response priming paradigm (Klotz & Neumann, 1999; Klotz & Wolff, 1995; Vorberg, Mattler, Heinecke, Schmidt, & Schwarzbach, 2003) was developed in the mid-1990's when the dominant theory in visual neuroscience was Goodale and Milner's (1992; Milner & Goodale, 1995) conception of ventral and dorsal pathways. Those pathways were thought to contain separate processing mechanisms for different types of input and output and to support qualitatively different types of representations, and so my natural starting point would be to expect different color representations underlying perception and action (even though I don't assume that Milner and Goodale's theory explains response priming). I'm not the only one to whom the idea of different representations seems natural: Breitmeyer, Ro et al. (2004) and Breitmeyer, Ro et al. (2007) explicitly conclude that conscious and unconscious color representations are the results of different transformations of the retinal cone contrast signals that put different emphases on certain wavelengths.

---

[1] A little primer on the terminology of color perception (based on Adelson, 2000). The light shining on a surface is an *illuminant*, which is characterized by its spectrum of electromagnetic radiation (energy as a function of wavelength). It hits upon a surface with a particular *reflectance profile*, which is the percentage of illuminant light reflected towards the observer, again as a function of wavelength. In the perception of gray shades, *brightness* is the perceived amount of light reflected from the surface, whereas *lightness* is the perceived reflectance of the surface. Lightness, but not brightness, is related to surface color. In the case of colored surfaces, those notions have to be generalized to functions of wavelength.

There are indeed reasons to believe that the color representations on which response priming is based might be quite different from the conscious perception of color. Conscious vision of surface color is not only phenomenally rich but also the result of complex information processing: To achieve color constancy, the visual system has to factor in such variables as number and location of light sources, the relative orientation of different surfaces in three-dimensional space, boundaries between shadow and light, invariances in color contrast across illumination boundaries, cross-reflections between surfaces, and sometimes even the time of day (because of the diurnal changes in the color of sunshine and skylight). Admirers of Rembrandt van Rijn, Jan Vermeer, or David Hockney will appreciate how difficult it is to capture the subtle interactions between lights and surfaces.

In stark contrast to all that, as far as we know now, response priming is based on relatively simple mechanisms of stimulus classification and rapid transmission through the visuomotor system. Converging evidence from lateralized readiness potentials (Eimer & Schlaghecken, 1998; Leuthold & Kopp, 1998; Vath & Schmidt, 2007; Verleger, Jaśkowski, Aydemir, van der Lubbe, & Groen, 2004), response hazards and response time distributions (Panis & Schmidt, 2016; Schmidt & Schmidt, 2014; Wolkersdorfer, Panis, & Schmidt, 2020), pointing trajectories (Brenner & Smeets, 2004; Schmidt, 2002; Schmidt & Schmidt, 2009), and force profiles (Schmidt, Weber, & Schmidt, 2014) indicates that the initial phase of the primed response is controlled exclusively by the prime: It controls the onset and the initial direction of the response, and the entire initial segment of response activation is independent of all properties of the target. This evidence led us to the theory that response priming is based on sequential feedforward sweeps through the visuomotor system (Lamme & Roelfsema, 2000), first one triggered by the prime and independent of the target, and only later one from the target as well (*Rapid Chase Theory*; Schmidt, Niehaus, & Nagel, 2006).

These findings rule out one concern of Skrzypulec's, namely that the prime color is classified only in relation to the visible target. He even argues that the priming effect may only occur if the system registers an identity of prime and target. But in pointing, force profile, and EEG studies, the prime clearly initiates the response before the target becomes available, and so the initial priming effect is not based on prime-target relations but on the prime alone. Note, however, that the necessary condition for this is only that the visuomotor system can *discriminate* red and green primes, not that it can identify them as belonging to (semantic?) categories of "red" or "green". But since this categorization is only implicit in the priming effect and could be performed on any representation, however sparse, that allows for a correct mapping of stimuli to responses, there is no reason why the representation underlying response priming shouldn't be relational. On the other hand, there is evidence that the processing carried by feedforward response priming can be quite complex, including discrimination of animals vs. plants (Haberkamp, Schmidt & Schmidt, 2013), illusory contours (Seydell-Greenwald & Schmidt, 2012), and Gestalt features of closure and symmetry (Schmidt & Schmidt, 2014).

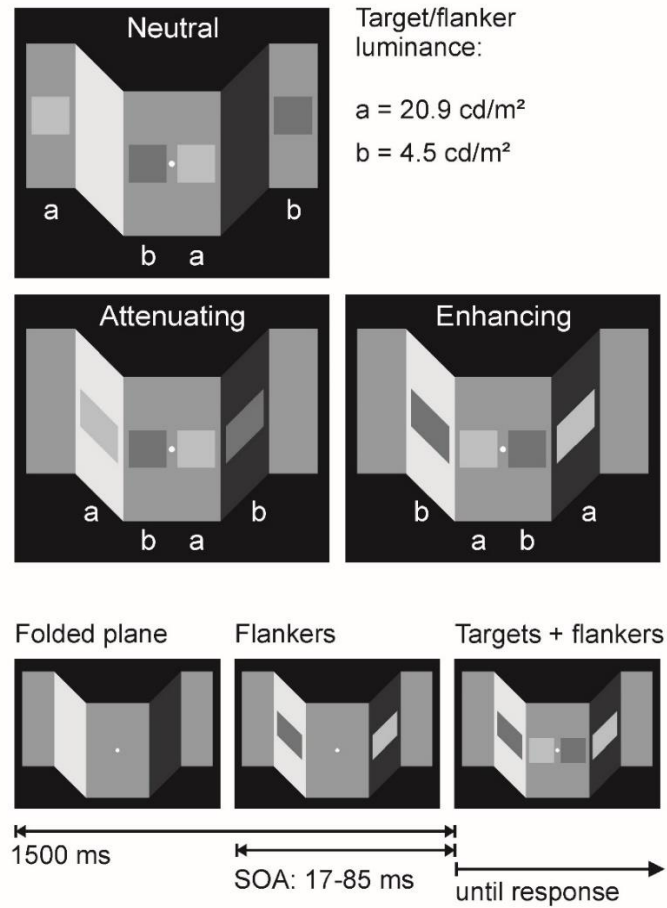

*Figure 1. Upper panel:* Two targets, one with high and one with low luminance, appeared in the central segment of a folded plane seemingly illuminated from the side. Two flankers, also one with high and one with low luminance, appeared either on the outer segments of the plane that were neutrally illuminated (neutral condition), or on the shadow and light segments. Depending on the illumination, the perceived brightness difference between the flankers could be attenuated or enhanced. Note that all stimuli labeled *a* or *b* have the same respective luminance; their different appearance results from the corrugated-plaid illusion. On consistent trials, the high-luminance prime appeared on the same side as the high-luminance target; on inconsistent trials (shown here), prime luminances were spatially switched compared to target luminances. *Lower panel:* Participants were instructed to first fixate on the center segment, then to respond by pressing a key on the side of the brighter target. Flankers appeared shortly before the targets; they were clearly visible and remained on screen until the response terminated the display. Figure reproduced from Schmidt et al. (2010).

Therefore, of course, Skrzypulec is right that color constancy is a crucial test. If color constancy can be demonstrated in response priming effects, the equivalence hypothesis is at least viable. However, there is direct experimental evidence clearly showing that response priming is not based on a color-constant representation, even if the primes are unmasked and shown for a prolonged time. Schmidt et al. (2010) used a version of the *corrugated plaid illusion* (Adelson, 2000) shown in Figure 1. Two targets, one brighter and one darker, appeared in the neutrally illuminated central segment, and the participants had to perform a speeded alternative-choice response on the side where the brighter target was. The primes were presented in a flanker arrangement; either in consistent or inconsistent configurations with respect to the targets. They were the same two shades of grey as the targets (therefore, one bright, one dark) and either appeared on the neutrally illuminated outer segments (neutral condition) or in the light and shadow zones. This creates a powerful illusion: When the bright prime appears in the light and the dark prime in the shadow, their apparent brightness difference is strongly attenuated; if the bright prime appears in the shadow and the dark prime in the light, their apparent brightness difference is strongly enhanced. Importantly, the flankers were unmasked and remained on screen until the participant responded, leaving all the time in the world to extract their surface colors.

*Table 1:* Results of Exp. 2 in Schmidt et al. (2010) where the contrast of the light and shadow zones was varied parametrically. In the enhancement conditions, local contrast was always regular, such that the more (less) luminant flanker was always more (less) luminant than the background. Brightness judgments were regular as well, such that the more luminant flanker was always judged brighter than the less luminant one. Priming effects were regular, such that responses were faster when flankers and targets were consistent in luminance. In the attenuation condition, the local contrast could be reversed compared to the flanker luminances, such that the more (less) luminant flanker could be darker (brighter) than the background. In all such cases, priming effects reversed as well, with faster responses now occurring in inconsistent. Brightness judgments, however, never reversed.

| **Enhancement:** | | | | | | | |
|---|---|---|---|---|---|---|---|
| light-shadow contrast: | 336.26 | 88.73 | 34.75 | 13.77 | 7.74 | 3.34 | 1.56 |
| local flanker contrast | regular | regular | regular | regular | regular | regular | regular |
| brightness judgments | regular | regular | regular | regular | regular | regular | regular |
| priming effects | regular | regular | regular | regular | regular | regular | regular |
| **Attenuation:** | | | | | | | |
| light-shadow contrast: | 336.26 | 88.73 | 34.75 | 13.77 | 7.74 | 3.34 | 1.56 |
| local flanker contrast | reversed | reversed | reversed | reversed | reversed | regular | regular |
| brightness judgments | regular | regular | regular | regular | regular | regular | regular |
| priming effects | reversed | reversed | reversed | reversed | reversed | regular | regular |

If response priming were based on a representation of surface color, we would expect the priming effect to follow the illusion: priming effects would be smaller in the attenuation condition and larger in the enhancement condition. But instead, the priming effect depended on local image contrast, not on the primes' perceived brightness. When we varied the amount of light and shadow in the scene, we found that priming effects *reversed* whenever the local contrast of the primes against their immediate background reversed (Table 1). This never happened in judgments of prime appearance: When participants were asked to adjust the targets in order to match the brightness of the flankers, their judgments did follow the perceptual illusion, but they always judged the more luminant prime to be brighter than the less luminant one, even if the local contrast switched. In other words, the more luminant prime could *look* darker but *prime* the response as if it was darker, and vice versa. This is direct evidence that brightness judgments are based on surface color while response priming is driven by local image contrast signals.

So Skrzypulec's warning that conscious and unconscious stimulus processing may be based on different representations, even qualitatively different ones, is entirely warranted. It constitutes a problem for those that claim that many if not most cognitive operations that are normally based on conscious cognition can also be performed unconsciously (most incautiously articulated by Hassin, 2013, without any regard for the methodological difficulties of many of the papers he reviewed; Schmidt, Haberkamp, & Schmidt, 2011; Shanks, 2017). Instead of speculating what all is possible for "the human unconscious", we should take the more modest approach of understanding a few of the many dissociations that can be established between, and among, measures of awareness and measures of implicit cognition (Irvine, 2017; Koster, Mattler, & Albrecht, 2020).

**References**

Adelson, E. H. (2000). Lightness perception and lightness illusions. In M. Gazzaniga (Ed.), *The visual neurosciences (2nd Edition)*, pp. 339-351. Cambridge, MA: MIT Press.

Biafora, M., & Schmidt, T. (2020). Induced dissociations: Opposite time-courses of priming and masking induced by custom-made mask-contrast functions. *Attention, Perception, & Psychophysics, 82*, 1333-1354.

Breitmeyer, B. G., Ro, T., Öğmen, H., & Todd, S. (2007). Unconscious, stimulus-dependent priming and conscious, percept-dependent priming with chromatic stimuli. *Perception & Psychophysics, 69*, 550-557.

Breitmeyer, B. G., Ro, T., & Singhal, N. S. (2004). Unconscious color priming occurs at stimulus- not percept-dependent levels of processing. *Psychological Science, 15*, 198-202.

Brenner, E., & Smeets, J. B. J. (2004). Color vision can contribute to fast corrections of arm movements. *Experimental Brain Research, 158*, 302-307.

Eimer, M., & Schlaghecken, F. (1998). Effects of masked stimuli on motor activation: Behavioral and electrophysiological evidence. *Journal of Experimental Psychology: Human Perception and Performance, 24*, 1737-1747.

Goodale, M. A., & Milner, A. D. (1992). Separate visual pathways for perception and action. *Trends in Neurosciences, 15*, 20-25.

Haberkamp, A., Schmidt, F., & Schmidt, T. (2013). Rapid visuomotor processing of phobic images in spider- and snake-fearful participants. *Acta Psychologica, 144*, 232-242.

Hassin, R. R. (2013): Yes it can: On the functional abilities of the human unconscious. *Perspectives on Psychological Science, 8*, 195-207.

Irvine, E. (2017). Explaining what? *Topoi, 36*, 95-106. https://doi.org/10.1007/s11245-014-9273-4

Klotz, W., & Neumann, O. (1999). Motor activation without conscious discrimination in metacontrast masking. *Journal of Experimental Psychology: Human Perception and Performance*, 25(4), 976–992.

Klotz, W., & Wolff, P. (1995). The effect of a masked stimulus on the response to the masking stimulus. *Psychological Research*, 58(2), 92–101.

Koster, N., Mattler, U., & Albrecht, T. (2020). Visual experience forms a multidimensional pattern that is not reducible to a single measure: Evidence from metacontrast masking. *Journal of Vision, 20(3):2,* 1-27.

Lamme, V. A. F., & Roelfsema, P. R. (2000). The distinct modes of vision offered by feedforward and recurrent processing. *Trends in Neurosciences, 23,* 571-579.

Leuthold, H., & Kopp, B. (1998). Mechanisms of priming by masked stimuli: Inferences from event-related brain potentials. *Psychological Science, 9,* 263-269.

Milner, A. D., & Goodale, M. A. (1995). *The visual brain in action.* Oxford: Oxford University Press.

Ro, T., Singhal, N. S., Breitmeyer, B. G., & Garcia, J. O. (2009). Unconscious processing of color and form in metacontrast masking. *Attention, Perception, & Psychophysics, 71,* 95-103.

Schmidt, F., Haberkamp, A., & Schmidt, T. (2011). Dos and don'ts in response priming research. *Advances in Cognitive Psychology, 7,* 120-131.

Schmidt, F., & Schmidt, T. (2010). Feature-based attention to unconscious shapes and colors. *Attention, Perception, & Psychophysics, 72(6),* 1480-1494.

Schmidt, F., & Schmidt, T. (2014). Rapid processing of closure and viewpoint-invariant symmetry: Behavioral criteria for feedforward processing. *Psychological Research, 78,* 37-54.

Schmidt, F., Weber, A., & Schmidt, T. (2014). Activation of response force by self-splitting objects: Where are the limits of feedforward Gestalt processing? *Journal of Vision, 14,* 1-16.

Schmidt, T. (2000). Visual perception without awareness: Priming responses by color. In T. Metzinger (Ed.), *Neural correlates of consciousness (pp. 157-179).* Cambridge: MIT Press.

Schmidt, T. (2002). The finger in flight: Real-time motor control by visually masked color stimuli. *Psychological Science, 13,* 112-118.

Schmidt, T., Miksch, S., Bulganin, L., Jäger, F., Lossin, F., Jochum, J., & Kohl, P. (2010). Response priming driven by local contrast, not subjective brightness. *Attention, Perception, & Psychophysics, 72,* 1556-1568.

Schmidt, T., Niehaus, S., & Nagel, A. (2006). Primes and targets in rapid chases: Tracing sequential waves of motor activation. *Behavioral Neuroscience, 120(5),* 1005–1016.

Schmidt, T., & Schmidt, F. (2009). Processing of natural images is feedforward: A simple behavioral test. *Attention, Perception, & Psychophysics, 71(3),* 594-606.

Schmidt, T., & Seydell, A. (2008). Visual attention amplifies response priming of pointing movements to color targets. *Perception & Psychophysics, 70,* 443-455.

Schmidt, T., & Vorberg, D. (2006). Criteria for unconscious cognition: Three types of dissociation. *Perception & Psychophysics, 68(3),* 489–504.

Seydell-Greenwald, A., & Schmidt, T. (2012). Rapid activation of motor responses by illusory contours. *Journal of Experimental Psychology: Human Perception & Performance, 38,* 1168-1182.

Shanks, D. R. (2017). Regressive research: The pitfalls of post hoc data selection in the study of unconscious mental processes. *Psychonomic Bulletin & Review, 24,* 752-775.

Skrzypulec, B. (2021). Contents of unconscious color perception. *Review of Philosophy and Psychology.* Published: 27 April 2021. https://doi.org/10.1007/s13164-021-00552-7

Tapia, E., Breitmeyer, B. G., & Shooner, C. R. (2010). Role of task-directed attention in nonconscious and conscious response priming by form and color. *Journal of Experimental Psychology: Human Perception and Performance, 36(1)*, 74-87.

Vath, N., & Schmidt, T. (2007). Tracing sequential waves of rapid visuomotor activation in lateralized readiness potentials. *Neuroscience, 145,* 197-208.

Verleger, R., Jaśkowski, P., Aydemir, A., van der Lubbe, R. H. J., & Groen, M. (2004). Qualitative differences between conscious and nonconscious processing? On inverse priming induced by masked arrows. *Journal of Experimental Psychology: General, 133,* 494-515.

Vorberg, D., Mattler, U., Heinecke, A., Schmidt, T., & Schwarzbach, J. (2003). Different time courses for visual perception and action priming. *Proceedings of the National Academy of Sciences USA, 100(10)*, 6275–6280.

Wolkersdorfer, M. P., Panis, S., & Schmidt, T. (2020). Temporal dynamics of sequential motor activation in a dual-prime paradigm: Insights from conditional accuracy and hazard functions. *Attention, Perception, & Psychophysics, 82*, 2581-2602.